\documentstyle[12pt]{article}
\newcommand{\be}{\begin{eqnarray}}
\newcommand{\ee}{\end{eqnarray}}
\newcommand{\nn}{\nonumber}
\newcommand{\ra}{\rightarrow}
\newcommand{\tl}{\tau^+{\tau}^-}
\newcommand{\bq}{b\bar{b}}
\newcommand{\bt}{b\bar{b}\tau^+{\tau^-}}
\begin{document}
\noindent IFT-96-03\\
hep-ph/9602292\\
\today

\vspace{1cm}
\begin{center}
{\large \bf  
Two-Higgs-doublet models
and the Yukawa process at 
LEP1  
  }\footnote{Supported in part by   
Polish Committee for Scientific Research and the EC grant under the contract 
CHRX-CT92-0004} \\

\vspace{1cm}
Jan Kalinowski and  Maria Krawczyk \\
Insitute of Theoretical Physics \\
Warsaw University \\ ul. Ho\.{z}a 69, 
00 681 Warsaw, 
Poland
\end{center}
\vspace{1cm}
\begin{abstract}
We investigate the  production of Higgs bosons $A/h$ via the Yukawa process 
$Z\ra f\bar{f} \ra f\bar{f} A/h $ taking into account  QCD corrections
and fermion mass effects. 
We estimate 
 the discovery reach of this process and/or bounds on the parameter 
space of two-Higgs doublet models which can be derived from 
the analysis of $b {\bar b} \tau^+ \tau^-$ events at LEP1.  
These limits have  important consequences for  extended models of 
electroweak interactions and their experimental verification/improvement
are welcome.
\end{abstract}
\vspace{1cm}

\section{Introduction}
Experimental searches at LEP1 for the neutral Higgs bosons in 
the context of extended 
models of electroweak interactions 
 have  been performed in the Bjorken ($Z \ra Z^*h$) and pair production
($Z\ra Ah$) processes. However, the Yukawa process ($Z\ra f\bar{f}\ra 
f\bar{f} A/h$), in which the Higgs boson is radiated off the final fermion, 
provides an alternative Higgs production mechanism. Although    
 the possible 
importance of this process in the context of two-Higgs doublet models, 
both in supersymmetric and non-supersymmetric versions, 
has been pointed out long time ago \cite{liter,nilles,zer}, 
experimentally it has not been looked for so far.

Recently there has been a renewed interest in the Yukawa process 
in the context of attempts to explain experimental measurements 
of $R_b$, the branching ratio for $Z$ decays to $b\bar{b}$ \cite{kane,inni}. 
In particular, in the paper \cite{kane} it has been  
argued,  based on (among others) 
a bound on parameter space $\tan\beta - M_A$ derived from  the Yukawa 
process $Z\ra \bq A$ followed by $A\ra \bq$ \cite{zer}, 
that problem of $R_b$ in the supersymmetric model could be explained 
only for $\tan\beta \sim 1$.   
This  bound  has been obtained from the 
simple assumption that if 20-30 $\bar{b}bA$ events had been
produced at LEP1 it is 
likely to have been noticed because the QCD background from
$Z\ra 4b$ is small. However, the analyses of $4b$-jet events, for 
example  in
$Z\ra hA\ra 4b$ searches \cite{aleph}, show that the background from $Z\ra
b\bar{b}gg$ due to misidentification of $b$-quark jets is much
more important. 
Although the Yukawa process has a different topology,
similar background problems may appear  in its experimental analysis
 and a definite conclusion can be
drawn after a dedicated experimental analysis in $4b$ channel will have been
performed. On the other hand, the $\bt$ channel is almost
background free and therefore it can be exploited to  estimate
limits on the parameters of the extension of the Standard Model. 

In  this paper we  estimate the 
parameter bounds and/or discovery reach of the Yukawa process 
using the $\bq\tl$ channel.
To this end we provide the simple formulae including all 
fermion mass terms for the matrix element
for process 
$e^+e^-\ra f\bar{f}\phi$ for both $\phi=h$ and $\phi=A$. 
They may prove useful in constructing Monte Carlo programs 
in order to include experimental constraints, efficiencies etc.

In the numerical analysis we also include QCD corrections which 
turn out to be very important.
Assuming that no signal events are found we obtain 
weaker bounds  on $\tan\beta - M_{\phi}$ than in \cite{kane}. 
Nevertheless our results demonstrate the physics potential
of the Yukawa process: even if the Higgs boson is not found  in this process, 
the exclusion plot derived from negative search will have interesting 
consequences for extensions of the Standard Model.

\section{Higgs particle production at LEP1}
In the minimal extensions of the Standard Model (SM), either  
supersymmetric (MSSM) or non-supersymmetric (2HDM), 
two  Higgs doublets are employed \cite{hunter}. 
In this paper we consider 2HDM  of 
type II in which, like in the MSSM,  one of the Higgs doublets 
couples to up-type, and the other to down-type fermions. 
After the symmetry breakdown 
 there are five physical Higgs particles: two CP-even neutral 
scalar Higgs bosons denoted
by $h$ and $H$ (we assume that $M_h\leq M_H$), 
a CP-odd neutral -- $A$ (usually called a pseudoscalar), and
a pair of charged particles $H^{\pm}$.
Their experimental discovery and measurement of their couplings  
provides  one of the
major motivations for and constitues a primary physics goal to be achieved
at present and future colliders.  

Currently the best experimental limits
on Higgs boson masses are obtained from LEP1 experiments performed at the 
$Z$ boson peak.  
At the tree level the neutral Higgs bosons $h$ and $A$ can be produced 
in $Z$ boson decays via 
three processes:
\be 
  Z & \rightarrow & Z^*h  \label{zh} \\
  Z & \rightarrow &  Ah   \label{ha} \\
  Z & \rightarrow & f {\bar f}\ra f {\bar f} A/h \label{yukawa}
\ee
In the SM the associated Higgs pair production process (\ref{ha}) 
is absent and the Yukawa process (\ref{yukawa}) is 
suppressed by 
a factor $(m_{f}/M_W)^2$ with respect to Higgs-strahlung process(\ref{zh}). 
Therefore the lower limit on the Higgs mass 
$M_{H_{SM}}> 65.2$ GeV has been derived \cite{mass} from  the negative 
search via the process (\ref{zh}) at LEP1.
  
In the two-doublet extensions of SM the Higgs-strahlung process (\ref{zh}) 
occurs at a lower rate
\be
\Gamma(Z\ra hZ^*)&=&\Gamma_{SM}(Z\ra H_{SM}Z^*) \sin^2(\beta-\alpha) 
\label{bjh}
\ee
because the $ZZh$ coupling is reduced by a factor $\sin^2(\beta-\alpha)$,  
where $\alpha$ and $\beta$ are the mixing angles in the neutral and 
charged Higgs sectors, respectively. The negative search in this process 
results in the upper limit \cite{sinlim} on $\sin^2(\beta-\alpha)$ 
for a given $M_h$, for example $\sin^2(\beta-\alpha)<0.1$ if $M_h<50$ GeV. 
However, for small $\sin^2(\beta-\alpha)$ one can  exploit  the Higgs pair 
production process (\ref{ha}), if kinematically allowed, because it is 
complementary to the process (\ref{zh}) in the sense that
\be 
\Gamma(Z\ra hA)=0.5 \Gamma(Z\ra \bar{\nu}\nu) \cos^2(\beta-\alpha)
\lambda^{3/2},
\ee
where $\lambda=(1-\kappa_h-\kappa_A)^2-4\kappa_h^2\kappa_A^2$, with 
$\kappa_i=m_i^2/M_Z^2$. 
Such a procedure
 has been adopted at LEP1. An independent contraint on the non-SM 
contributions to $Z$ decay 
 is  obtained from the analysis of the $Z$-lineshape, where the process (5)
 also contributes.

In this paper we consider a scenario in which the
process (\ref{zh}) 
is suppressed (small $\sin^2(\beta-\alpha)$) {\it and} the process 
(\ref{ha}) is closed kinematically ($m_h+m_A >m_Z$). Then the Yukawa 
process (3)  becomes a dominant Higgs  production mechanism at LEP1. 
Such a scenario can easily be realized in the 2HDM where 
 the Higgs masses and mixing angles are not
constrained theoretically. On the other hand,  
 in the MSSM, due to supersymmetry
relations, the Yukawa process may become dominant only 
for $m_h>50$ GeV and $\tan\beta>10$ \cite{nilles}. 

The number of expected events crucially  depends on $h\bar{f}f$ 
and $A\bar{f}f$ couplings: the SM 
Higgs coupling $(\sqrt{2}G_F)^{1/2} m_{f}$ is multiplied by the 
factors $g_{\phi f\bar{f}}$ given in  Table 1, 
where $D$ ($U)$ is a generic notation  for down-type (up-type) fermions, 
respectively. 

\begin{center}
\begin{tabular}{|c|c|c|}\hline
$g_{\phi f\bar{f}}$ &     $f=D$                & $f=U$ \\ \hline
$\phi=h$           &  $-\sin\alpha/\cos\beta$ & $\cos\alpha/\sin\beta$ \\ 
 $\phi=A$          &    $ \tan\beta$          & $1/\tan\beta $ \\ \hline 
\end{tabular}
\end{center}

\begin{center} Table 1: {\it Higgs boson couplings to fermions relative to the 
SM Higgs couplings.}
\end{center}

\vskip 1mm 
In the interesting large $\tan\beta$ case the couplings of 
{\it both} $A$ and $h$  to 
down-type fermions are strongly enhanced by the same factor $\tan\beta$ 
because small $\sin^2(\beta-\alpha)$ implies\footnote{A detailed 
discussion will be given elsewhere  \cite{dlakr}. In what follows we take $\alpha=\beta$
for the $h\bq$ and $h\tau \tau$ couplings.},  
to a good approximation, that  
$\sin\alpha/\cos\beta \sim 
\tan\beta$.
In such a case the Yukawa process can be a source of a significant number of 
Higgs bosons $h$ or $A$, whichever is lighter. 
For example, at LEP 1 one can expect \cite{lekki} 
for $M_{h/A}=10$ GeV and $\tan\beta=20$ 
a few thousand of $bbh$ or $bbA$ events 
in 10$^7$ $Z$ decays 
which  corresponds to the $Z$ partial decay width of $\sim$ 1 MeV. 

\section{The Yukawa process}

The rate for the process is very sensitive to the
fermion mass effects both in matrix elements and in the kinematical 
limits as demonstrated in our previous paper \cite{lekki}.
The analytical formulae for the matrix element 
of the process  $e^+e^-\ra f {\bar f} \phi$, 
including all fermion mass terms, can be written in a compact 
form. We present them in Appendix, as 
 they may appear to be useful in constructing
Monte Carlo programms in order to include experimental effects.

In this paper we consider the effect of QCD corrections in the 
Yukawa process.
In the case when the Higgs boson is radiated off the $b$ quark ($f=b$)
and in hadronic decays ($\phi\ra b\bar{b}$) the QCD 
corrections are included by using the running 
quark mass $m_f=m_b(M_{\phi}) $ in the Higgs-quark coupling. 
This has an important effect on the production cross section. For 
example, for $M_{\phi}=50$ GeV the cross section is reduced by a 
factor 2 in comparison to fixed quark mass case, $i.e.$ neglecting 
QCD corrections.

We observe that the cross section for the Yukawa process is
proportional to $\tan^2 \beta$. The additional dependence
on $\tan\beta$ that appears in the branching ratios for dominant
Higgs decay modes $\phi\ra \tau^+{\tau}^-$ and $b\bar{b}$ for
$\tan\beta>5$ is very weak and therefore can be neglected{\footnote
{ We use the program developed in ref.\cite{dkz} to calculate the 
Higgs decay branching ratios.}}. 
To derive the
limits on $\tan\beta-M_{\phi}$ 
(with $\phi=h,A$) we proceed as follows:
\begin{enumerate} 
\item[a)] We perform the
calculations at the $Z$ pole for 10$^7$ $Z$'s produced.
\item[b)] We consider two different final state configurations: 
$\bq(\tl)$ and $\tl(\bq)$, 
where the first pair denotes the fermions 
produced in $e^+e^-$ collisions which radiate the Higgs boson 
(either $h$ or $A$) and the pair in the parenthesis is coming from 
Higgs decay. 
\item[c)] As we discussed, the $\bq\tl$ channel 
seems to be the most promising both from theoretical (negligible 
 QCD background) 
and experimental (relatively good efficiency) points of view. 
For example, the L3 Collaboration \cite{L3}
quotes the acceptance for signal events from $Z\ra hA\ra \bq
\tl$ of the order of 5\%.     
We assume that in the analysis of the Yukawa process similar efficiency 
can be achieved 
\item[d)] We assume no signal
events and we derive the bounds corresponding to  95\% CL.
\end{enumerate}

The results are presented in Fig.1 for the pseudoscalar Higgs boson $A$ and 
in Fig.2 for the scalar $h$.  
Parameters above the lines are excluded. 
The solid lines represent the limits obtained with running $b$-quark
mass $m_b(M_{\phi})$, and the dashed lines correspond to the 
case of fixed $b$-quark
mass $m_b=4.6$ GeV, $i.e.$ neglecting QCD corrections. The QCD corrections
lower the Higgs-quark coupling and consequently weaker bounds
can be established. 
For comparison  the 
dotted lines delineate the excluded region based on 20 $4b$-jet 
events expected.  

Note that in the MSSM  case the mass range of interest  for $M_h$ and 
$M_A$ is lying
 above, say 50 GeV.
On the other hand in the 2HDM  relatively light Higgs bosons
 are still not excluded, for 
example for $\tan\beta=20$ the Higgs masses as low as  25 GeV, and for 
$\tan\beta=55$ masses of 45 GeV are not excluded.

To summarise, the Yukawa process provides interesting limits on
the parameter space of 2HDM and MSSM. 
Both QCD and fermion mass effects influence the results
considerably.

 Although the first attempt to include the Yukawa process in Higgs
searches at LEP1 has been made recently by L3 
Collaboration \cite{L3}  nevertheless 
this process still awaits a serious experimental investigation.

\vspace{1cm}
\noindent {\large\bf Acknowledgments}\\
We are grateful to Andre Sopczak for his interest in this subject
and comments
on the experimental analysis performed at L3.
We would like to thank    P. Chankowski, C. Wagner and  P. Zerwas
 for useful discussions.

\bigskip
\noindent { \Large \bf Appendix}

\vskip 2mm 
\noindent We consider  the Yukawa process   $$e^-(k_1)e^+(k_2)
\rightarrow f(p_1) {\bar f}(p_2) \phi(l)$$  
 with $f=b$ or $\tau$ and the corresponding 4-momenta are given in brackets. 
The electron mass is neglected ($k_{1,2}^2=m_e^2=0$) and
 the final state fermion (pole) mass 
is denoted by $m$ ($p_{1,2}^2=m^2$).
The Higgs mass is denoted by $M_{\phi}$ and $s=(k_1+k_2)^2$. 

The differential cross section $ E_1E_2\mbox{d}\sigma/\mbox{d}^3p_1 
\, \mbox{d}^3p_2$ reads  
\be
E_1E_2{{\mbox{d}\sigma}\over {\mbox{d}^3p_1 \, \mbox{d}^3p_2}}=
{{1}\over{128  \pi^5 s^2 x_{\phi}}}
{\overline {|{\cal M}|}}^2 \delta(x_{\phi}-2+x_1+x_2)
\ee
 \be
x_1={{2 E_1}\over{\sqrt{s}}}, \;\; x_2={{2 E_2}\over{\sqrt{s}}}, \;\;
x_{\phi}={{2 E_{\phi}}\over{\sqrt{s}}},
\ee
where 
 $E_{1,2}$ and $E_{\phi}  $
 denote CM energies of fermions and the Higgs boson $\phi$, respectively.
  
The matrix element, averaged over the initial and summed over final 
fermion polarizations,   
can be cast in the following form
\be
{\overline{ |{\cal M}|}}^2&=&
64 \pi^2 \alpha^2_{el}\, N_c\, {{G_F}\over{\sqrt{2}}}\,
 g^2_{\phi f {\bar f}}\, m_f^2\, s  \nn \\
&&\{Q_e^2 Q_f^2\, {{\gamma^2}\over{s^2}}
 +{{(a_e^2+v_e^2)[ v_f^2\gamma^2+
a_f^2 \zeta^2]+2 a_e v_e a_f v_f\chi}\over{(s-M_Z^2)^2+M_Z^2\Gamma_Z^2}}\nn \\
&&+Q_e Q_f {{(2 v_e v_f\gamma^2 + a_e a_f \chi)(s-M_Z^2)}
\over{s((s-M_Z^2)^2+M_Z^2\Gamma_Z^2)}}\},
\ee
The color factor is $N_c=3$ for  $f=b$ and $N_c=1$ for $f=\tau$.
We distinguish  the fermion mass which enters the coupling, $m_f$, from 
the pole mass $m$. In the case of the $\tau$ lepton we take $m_f=m=m_{\tau}$.
However, in the case of the $b$-quark, we include 
 QCD corrections by taking the running quark mass $m_f=m_b(M_{\phi})$. 
$Q_e$,$Q_f$ are electric charges of electrons and fermions, respectively 
(in units of $e$) whereas
$a_e(a_f)$,$v_e(v_e)$ correspond to axial and vector couplings
of electron (fermion $f$) to $Z$-boson with 
$a=I_{3L}/2\sin\theta_W\cos\theta_W$ and $v=(I_{3L}-2Q\sin^2\theta_W)
/2\sin\theta_W\cos\theta_W$.

 The contributions of the   photon, $Z$-boson exchange and the $\gamma-Z$ 
interference 
are denoted by $\gamma^2$, $\zeta^2$ and $\chi$,
respectively, and can be written as follows:

\noindent a){\it Pseudoscalar}\\
In the case of the pseudoscalar production, $\phi=A$,  we have 
$g^2_{A f\bar{f}}=\tan^2\beta$, $M_{\phi}=M_A$ and  
\be
\gamma^2&=&{{1}\over{\xi_1^2\xi_2^2}}{{M_A^2}\over{s}}
 [x_A^2 {{-2m^2-s}\over{s}}+
 x_A(\Pi_1\xi_2+\Pi_2\xi_1)]     
 +{{1}\over{\xi_1\xi_2}}
[x_A^2 -
2( {{M_A^2}\over{s}}+1) x_{T}^2]\nn \\
\zeta^2&=&{{1}\over{\xi_1^2\xi_2^2}}{{M_A^2}\over{s}}[x_{21}^2 
{{2m^2-s}\over{s}}-
x_{21}(\Pi_1\xi_2-\Pi_2\xi_1)]     \nn \\
&&+{{1}\over{\xi_1\xi_2}}[x_{21}^2 + x_{21}  \Pi_- 
+ 2( {{M_A^2-4 m^2}\over{s}}+1-x_A)x_{T}^2] \nn \\
\chi&=&{{1}\over{\xi_1^2\xi_2^2}}{{M_A^2}\over{s}}
[\Xi_1 \xi_2^2-\Xi_2 \xi_1^2] 
-{{1}\over{\xi_1\xi_2}}[
       (\Xi_1 \xi_2-\Xi_2 \xi_1)-x_A \Delta\Pi] 
\ee

\noindent b) {\it Scalar}\\
For the  scalar, $\phi=h$, we have $g^2_{h f\bar{f}}=(\cos\alpha/\sin\beta)^2$,
$M{\phi}=M_h$ and 
\be
\gamma^2&=&{{1}\over{\xi_1^2\xi_2^2}}{{M_h^2-4m^2}\over{s}}
[x_h^2 {{-2m^2-s}\over{s}}+x_h(\Pi_1\xi_2+\Pi_2\xi_1)] \nn \\
&&+{{1}\over{\xi_1\xi_2}} [x_h^2 -  2( {{M_h^2-4m^2}\over{s}}+1)x_{T}^2]\nn \\
\zeta^2&=&{{1}\over{\xi_1^2\xi_2^2}}\{ {{M_h^2}\over{s}}
[x_{21}^2 {{2m^2-s}\over{s}}- x_{21}(\Pi_1\xi_2-\Pi_2\xi_1)]\nn \\ 
&&-{{4m^2}\over{s}} [x_{h}^2 {{2m^2-s}\over{s}}+
x_{h}(\Pi_1\xi_2+\Pi_2\xi_1)]\} \nn \\ 
&&+{{1}\over{\xi_1\xi_2}}[x_{21}^2 + x_{21}  \Pi_- 
-16 {{m^2}\over{s}}(x_h- {{ M_h^2}\over{s}})
+ 2( {{M_h^2}\over{s}}+1- x_h) x_{T}^2] \nn \\
\chi&=&{{1}\over{\xi_1^2\xi_2^2}}[{{M_h^2}\over{s}}
(\Xi_1 \xi_2^2-\Xi_2 \xi_1^2)
-{{4 m^2}\over{s}} x_h (\Xi_1 \xi_2-\Xi_2 \xi_1)]      \nn \\ 
&&-{{1}\over{\xi_1\xi_2}}[ (\Xi_1 \xi_2-\Xi_2 \xi_1)-x_h \Delta\Pi] 
\ee
where 
the following variables are introduced
\be
&&x_{21}=x_{2} - x_{1}   \\
&&\xi_1=1-x_1, \;\; \xi_2=1-x_2   \\
&&x_{T}^2=4(p_1+p_2)\cdot k_1\, (p_1+p_2)\cdot k_2/s^2-1+x_h-M_{\phi}^2/s \\
&&\Pi_1=8 (p_1\cdot k_1\, p_1\cdot k_2)/s^2, \;\;\Pi_2=8 (p_2\cdot
k_1\, p_2\cdot k_2)/s^2\\ 
&&\Pi_-=\Pi_1-\Pi_2\\
&&\Xi_1=4 p_1\cdot (k_1-k_2)/s, \;\; \Xi_2=4 p_2\cdot
(k_1-k_2)/s \\ 
&&\Delta\Pi=8 (p_1\cdot k_1\, p_2\cdot k_2-p_1\cdot k_2\, p_2\cdot k_1)/s^2
\ee

\newpage

\begin{figure}[t]
\vskip 7in
\includegraphics{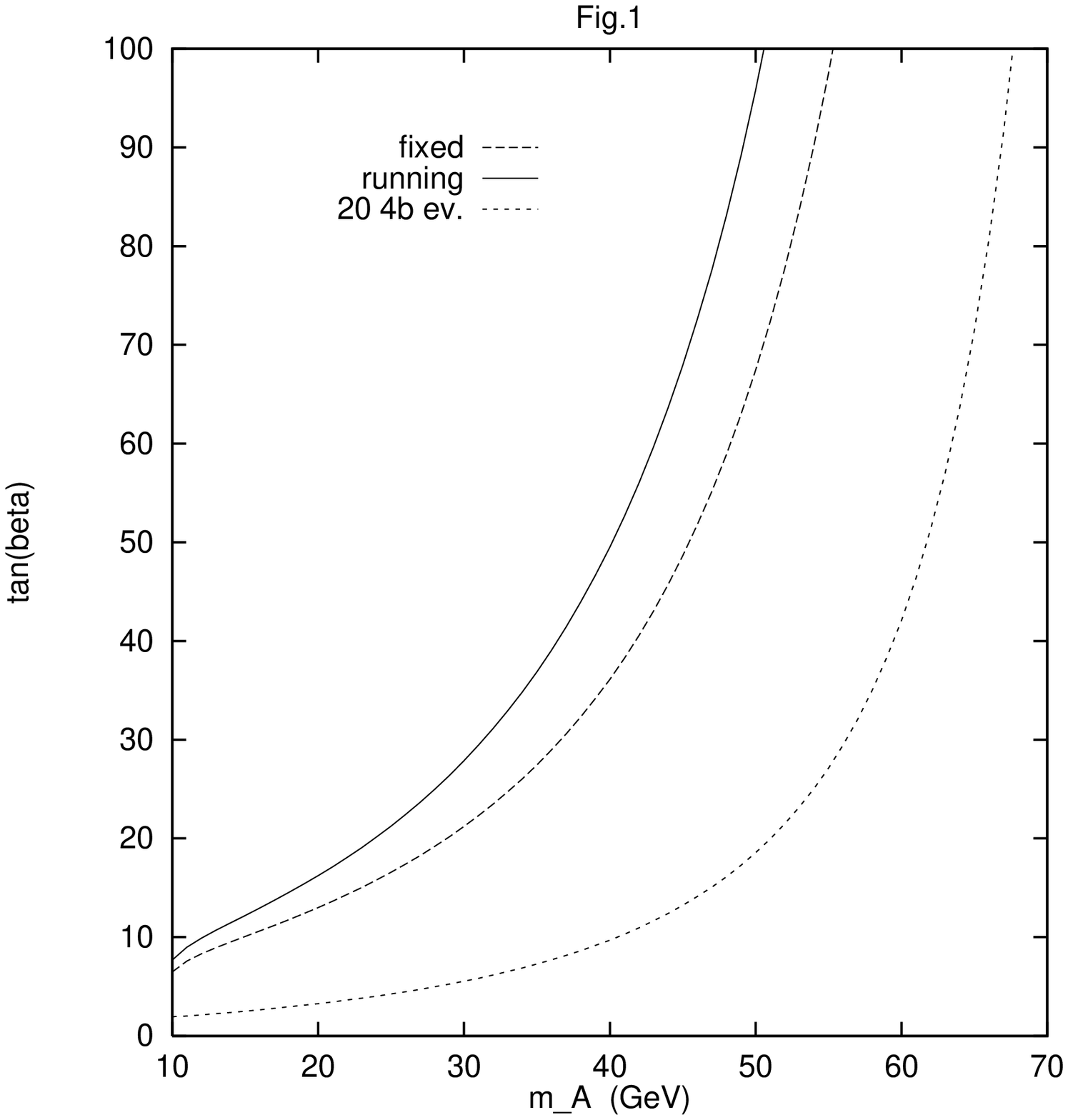}
\caption{The $\tan\beta - M_A$ exclusion/discovery  plot for 
the Yukawa process for $A$ production.
The region above the lines is excluded if no signal in $\bt$ mode
is observed. The solid line represents the results 
when QCD corrections are accounted for by taking the running $b$-quark mass 
and   the dashed line corresponds to fixed $b$-quark mass $m_b=4.6$ GeV. 
For comparison 
the dotted line shows the region that can be excluded based on 20 events 
with $b$-quark jets (using the running $b$-quark mass).}   
\end{figure}
\clearpage

\begin{figure}[t]
\vskip 7in
\includegraphics{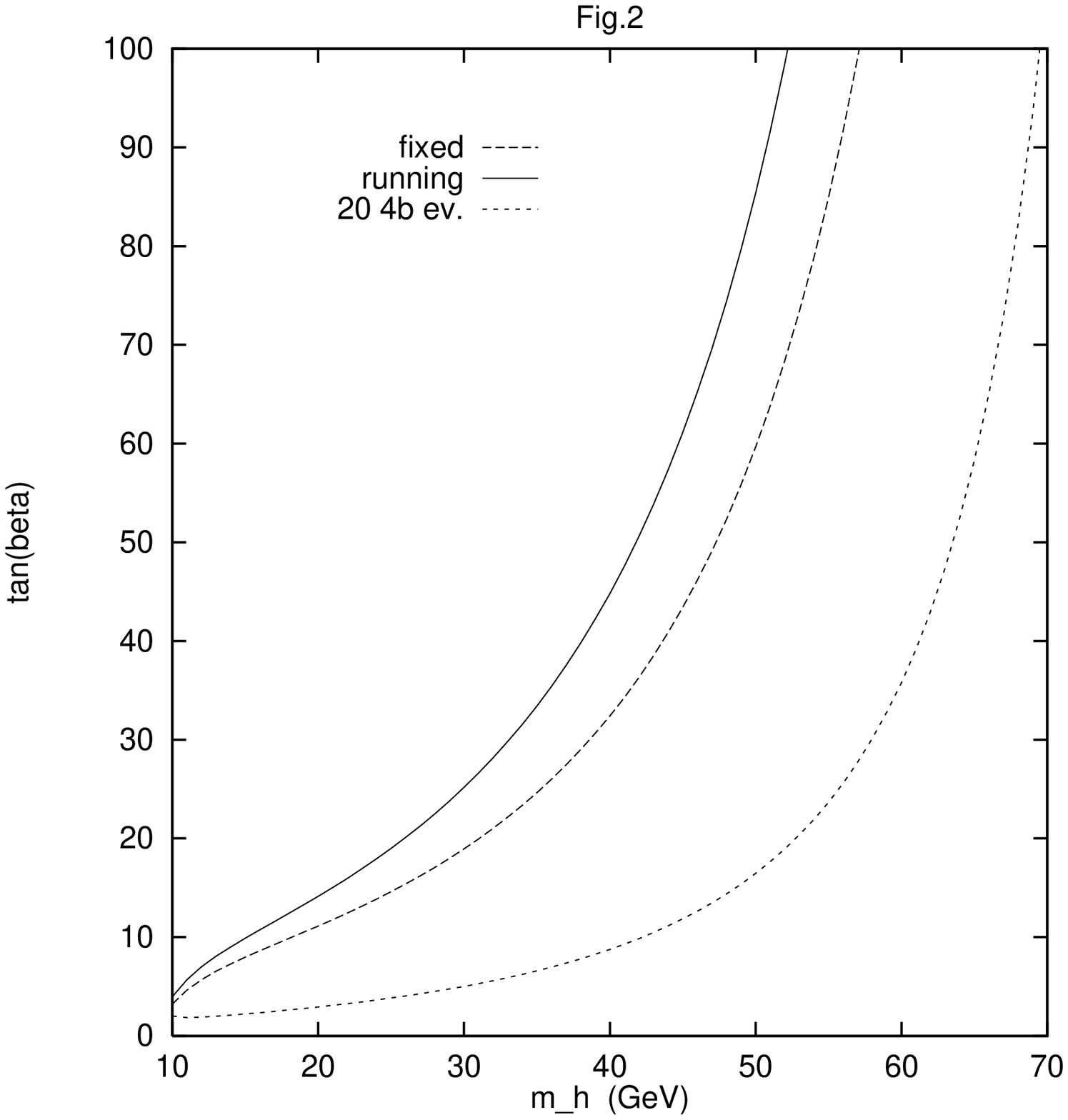}
\caption{The $\tan\beta - M_h$ exclusion/discovery  plot for 
the Yukawa process for $h$ production.
The region above the lines is excluded if no signal in $\bt$ mode
is observed. The solid line represents the results 
when QCD corrections are accounted for by taking the running $b$-quark mass 
and   the dashed line corresponds to fixed $b$-quark mass $m_b=4.6$ GeV. 
For comparison 
the dotted line shows the region that can be excluded based on 20 events 
with $b$-quark jets (using the running $b$-quark mass).}
\end{figure}
\clearpage

\end{document}